# Seeing and interacting with the invisible: A powerful tool for the learning of science

*Vera Montalbano*
Department of Physical Sciences, Earth And Environment, University of Siena, Italy.

*Abstract:* In order to describe natural phenomena, science develops sophisticated models that use mathematical and formal languages which seem, and often are, very far from common experience. When a phenomenon is not accessible to our senses, its description is indirect and understanding can be difficult for those who are not trained to understand the consequences of formal languages used by scientists. When one succeed to obtain a direct visualization of a phenomenon inaccessible to senses, it is possible to get a deeper understanding since a very effective channel of learning is involved. A wider and more profound result in learning process can be obtained if the physical system utilized for visualization enables direct interaction with the phenomenon. From the infrared vision to cosmic rays, from the magnetic field to the flow of energy, many phenomena can be suitable for building systems that allow capturing a greater awareness of the physical world. Some examples of such systems are given for relevant topics in physics and for mathematical tools. They were designed for a summer school for students in the last years of high school or for deepening laboratories addressed to talented students in secondary school, but with some attention it is possible to adapt them to other cases like high school classes or undergraduate students.

# INTRODUCTION

Modelling plays a major role in scientific methodology and in providing an explanation of some aspect of natural phenomena. A model gives a mental representation of a phenomenon that can be visualized and shared with other people. Since one of main purpose of science education is to disseminate the outcomes of science, modelling is crucial in learning process (Gilbert, Boulter, & Elmer, 2000). Visualization is an essential and unavoidable aspect of modelling for a deep understanding of the relationship between the model and the phenomenon which it describes (Gilbert & Jones, 2010).

Many topics in physics, as well as in other sciences, are inaccessible to our senses and usually are modellized by using mental models. Sometimes however, it is possible to create a model that allows a visualization of qualitatively and/or quantitatively relevant issues in a direct way. Human knowledge is more profound and immediate if it passes through the use of the senses. Among these, the sight is the most important and utilized in today's society. Thus, if a model allows a direct visualization of invisible aspects of nature, the learning process ca be easier and faster. In addition, the directness of the model can become a tool for analysis of some features of the phenomenon and suggest relationships and unexpected behaviour allowing students to have a full experience of the use of models in science.

Models used for complex topics often can be introduced and studied by a powerful tool of modern science, i.e. the simulation. Even in science education this instrument is used mainly because cheaper, but teaching with computerized simulations is not straightforward (Murnane, 2002). Frequently it remains open the question of how the results of the simulation are connected to reality and also when the model is simple in its implementation, students may have the feeling of moving in an artificial context because they have not enough technical competencies for distinguishing simulation artefacts from genuine features of the model.

How do physicists use external visualization (Gilbert, 2010) in research?

The use of visualization in physics has been analyzed. The focus is put on the use of visualization in modelling of complex phenomena and three different levels have been identified in physics research.

Is it possible to improve the effectiveness of physics education by using the more unusual of them? Which activities are useful in classroom practice?

Some attempt of investigating these issues have been realized within the National Plan for Science Degree (Montalbano, 2012; Sassi, Chiefari, Lombardi & Testa, 2012) in a summer school of physics (Benedetti, Mariotti, Montalbano, & Porri, 2011; Montalbano & Mariotti, 2013) and with deepening laboratory with few interested students (Montalbano & Di Renzone, 2012). The pilot study was qualitative and involved small groups of students from high school (15 - 18 years) in different times and situations. Examples in the following topics have been examined: magnetic field and mechanical systems, energy transport and transformations.

# HOW PHYSICISTS USE VISUALIZATION

The first step was to analyze the external representations that physicists use in the research. Mainly, it is possible to identify three different types of images that are described below. In Table 1, the main characteristics of each type of external representation, usually utilized in physics, are summarized. Moreover, some example for every use is indicated between figures in the following.

Table 1

*Different types of external representations used in physics in scientific research.*

| Type | Characteristics | Examples |
| --- | --- | --- |
| A<br><br>Images for summarizing a model | Theor & Exp info<br>Not in scale<br>Not interactive<br>No direct connection with a physical system | fig. 1, fig. 2,<br>fig.4 (upper part)<br>fig. 5 (first on the left) |
| B<br><br>Images from computer simulation | Can be in scale<br>Can be interactive<br>Theor hypoth & data<br>Connection is indirect and depends on theor hyp & data | fig. 3,<br>fig. 5 (middle one) |
| C<br><br>Physical systems with a direct and interactive visualization of a phenomenon | Visualization depends very weakly from hypothesis<br>In scale<br>Interactive<br>Direct connection with physical system | fig. 4 (one below)<br>fig. 5 (on the right) |

## Images for summarizing a model

In these kind of images, much information is summarized, both theoretical and experimental, but usually only in a qualitative way.

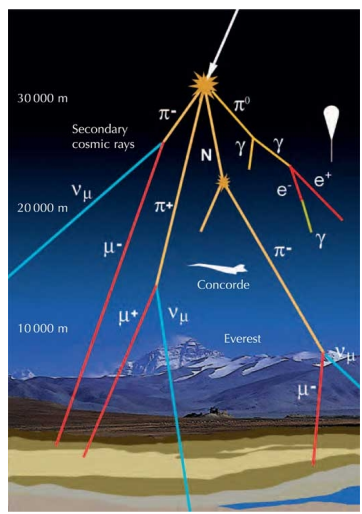

*Figure 1. A schematic representations of cosmic rays (LNL website)*

For example in fig. 1, cosmic rays are represented not in scale (since particles in the showers have ultra relativistic velocity, all angles are much smaller than the ones drawn; balloon experiment and Concorde are not in scale with mountains and so on).

In fig. 2, atoms are trapped by coherent light from crossed laser beams. There is no way of select five or more atoms, of course, this is only a visual way to show an ordered trapping in space. Moreover, spatial and temporal coherence proper of a laser beam are represented by mean of red rough strips which are in some sense remotely reminiscent of the wave fronts of the electromagnetic potential.

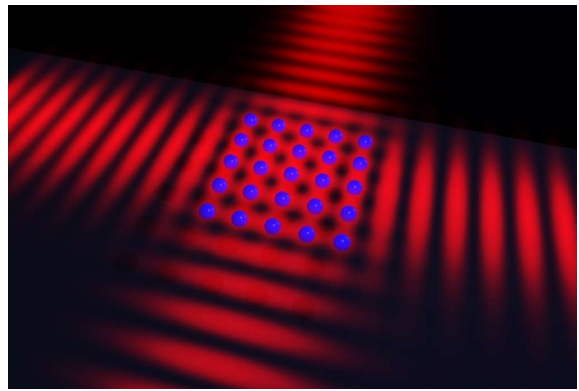

*Figure 2*. Ultracold atoms trapped in optical lattices (Inguscio, 2012).

These visualizations are no direct connected to a physical system but are useful for summarising a model, outline which aspects are really important and which ones can be neglected.

## Images from computer simulation

An impressive number of images are produced by simulations like the one showed in fig. 3 for a cosmic ray shower. They are very useful in research for testing hypotheses and systems before a real proof in experiment. Computer graphics allows to interact with this images and from this point of view they can be utilized in teaching and learning science.

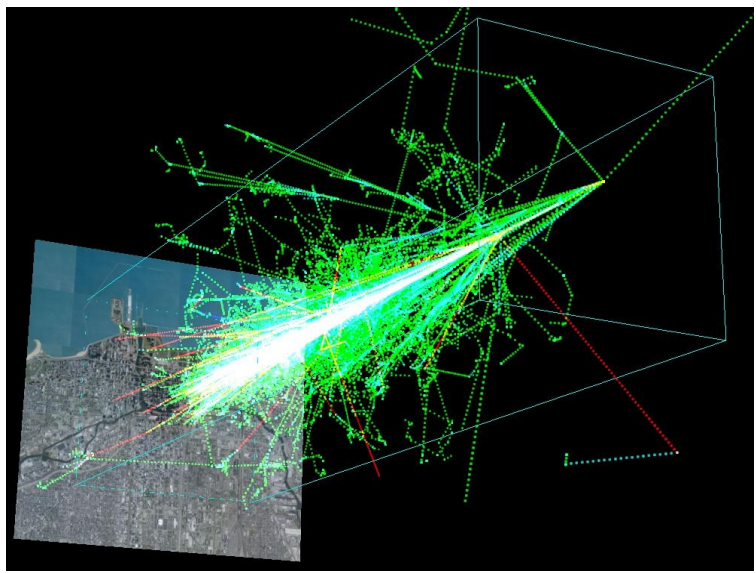

*Figure 3*. A cosmic ray shower simulations (AIRES Cosmic Ray Showers website).

The figure 3, for example, can be turned in 3D in every direction in such a way that one can be aware of real cosmic ray's spatial distribution comparing images with different perspectives with the city's map on the ground.

On the other side, students have a wide experience of virtual worlds, and often identify simulation with a tool that can be used for create situations very far from the real world. Thus, usually simulations are not very effective in classroom practice.

## Direct visualization of invisible phenomenon

The last type of visualization is performed by choosing a physical system that allow a direct representation of the phenomenon, i.e. in scale and any change in the parameters of the phenomenon corresponds to a change in the system. Thus, we obtain a simulation realized by mean of a connect physical system. In this case the visualization have no dependence from hypothesis and model that we are using for describing the phenomenon.

Therefore, the external representation becomes a useful tool for studying a physical process in a quantitative way. For example in fig. 4, optical trapping is summarized in the upper images with different boundary conditions. In the lower ones, a systems of ultra cold atoms is trapped in different way showing the behavior predicted by the model of the phenomenon.

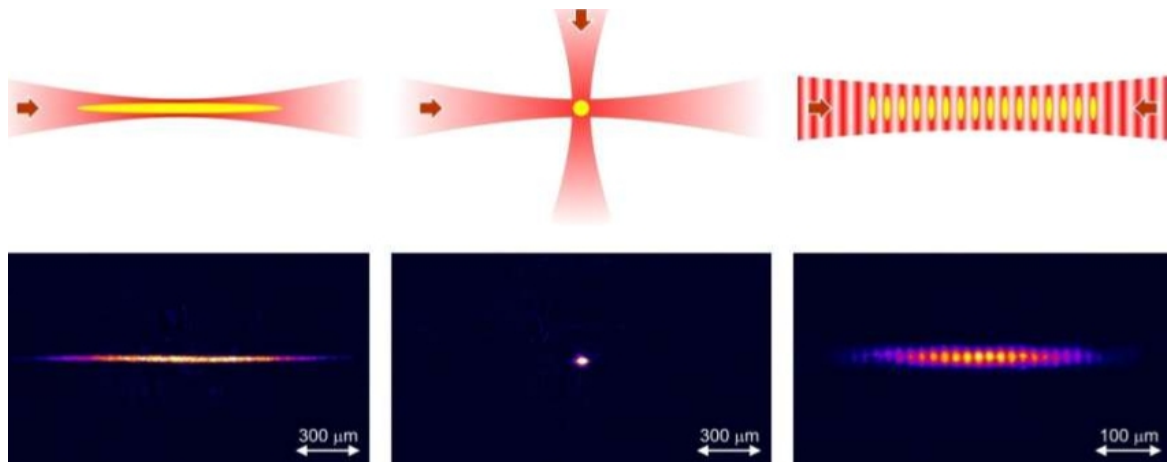

*Figure 4*. Optical trapping: single-beam trap, crossed-beam trap, optical lattice ; the model (upper) and a physical realization by mean of trapped ultracold atoms. The images are formed by the photoemission in visible light (false color indicates light intensity) and allows to measure directly how many atoms are trapped (Inguscio, 2012).

Another example is given in fig. 5, where it is showed the Anderson localization on a lattice surface. In this case, a quantum behavior was foreseen, than investigated by simulation and finally realized in laboratory.

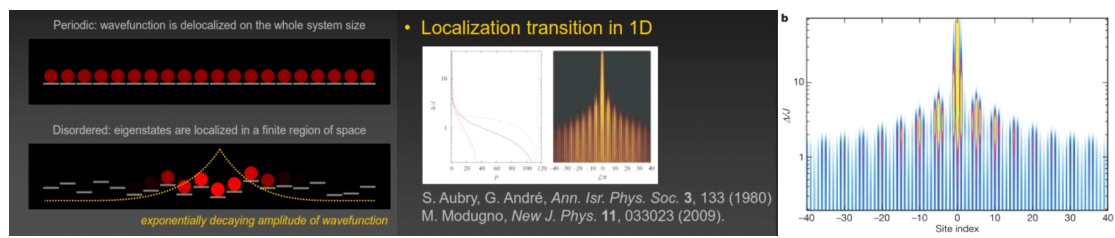

*Figure 5*. Anderson localization: schematic model, simulation and realization in laboratory visualized by an ultra cold gas (Inguscio, 2012).

# EXAMPLES IN PHYSICS EDUCATION

In order to clarify how visualization in physics lab can improve physics understanding of complex phenomena, some physical systems were selected. Let us start with a very known and versatile system: the oscilloscope.

## A "trivial" example: the oscilloscope

A physical quantity (the electric potential V) is measured at different times and, by means of a trigger, it is visualized as a function of time. The oscilloscope allows to visualize directly a wide range of physical quantities in function on time for many systems (electric, electro mechanics, acoustics and so on) as soon as one physical quantity, relevant for the description of the system, can be transduced into an electric signal. Many research can be performed by using this device but it is widely utilized in physics education too.

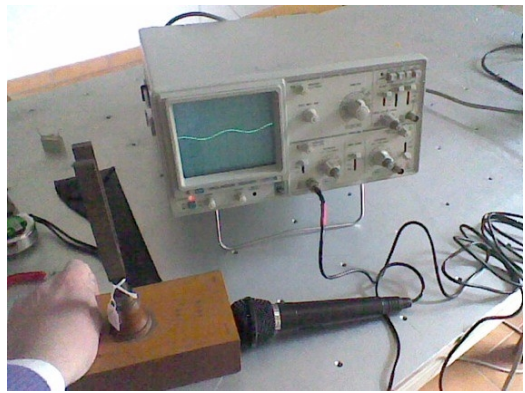

*Figure 6.* On the display of oscilloscope the sound wave produces, by mean of a sensor (a microphone), a signal that reproduces directly many wave features which can be easily measured, such as frequency, period, intensity.

In fig. 6 it is shown like example a set-up for direct visualization in the case of sound waves. In this way it is possible to study a pure sound or superposition of pure sound (beats).

## Magnetic field

### *On showing magnetic field by using magnetic flux*

In physics education in order to construct a model, an important role for visualization can be played for all phenomena which are not visible in the sense that we have no biological sensor for detecting physical quantities that are relevant. This is the case for magnetic or electric field.

In fig. 7, it is shown a sequence of pictures in which a magnet is falling inside a metal tube. Eddy currents are produced by induction and the magnet is slowed in its falling. The visualization of the magnet during the falling through a metal surface is always very effective with students. The next step is to outline that the change in the green sensor corresponds to variation of magnetic flux and then can be connected to Faraday's law.

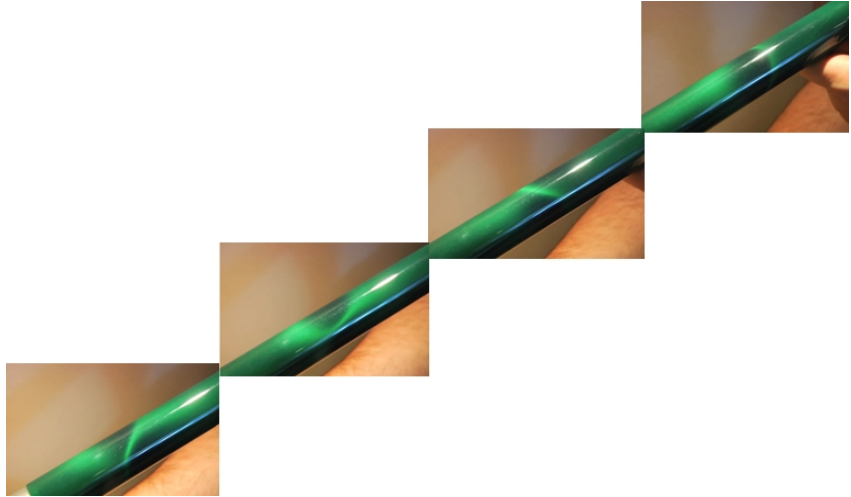

*Figure 7.* The sequence allows to visualize a permanent magnet rolling and falling in a metal tube by using magnetic field flux variation.

## On showing magnetic field by using ferrofluid

Another very effective tool in electromagnetism is ferrofluid. This liquid is a colloidal solution of ferromagnetic nanoparticles which have a superparamagnetic behavior. Thus, in presence of a magnetic field the liquid changes its surface in order to follow the magnetic field lines. In many case this characteristic can be used in a learning path on electromagnetism.

**Spatial distribution.** The more impressive use of ferrofluid is with ceramic magnet (more strong than usual ones). In fig. 8, the spatial distribution of magnetic field around magnet is shown. It is important to underline that this system allows to visualize magnetic field lines in 3 dimensions.

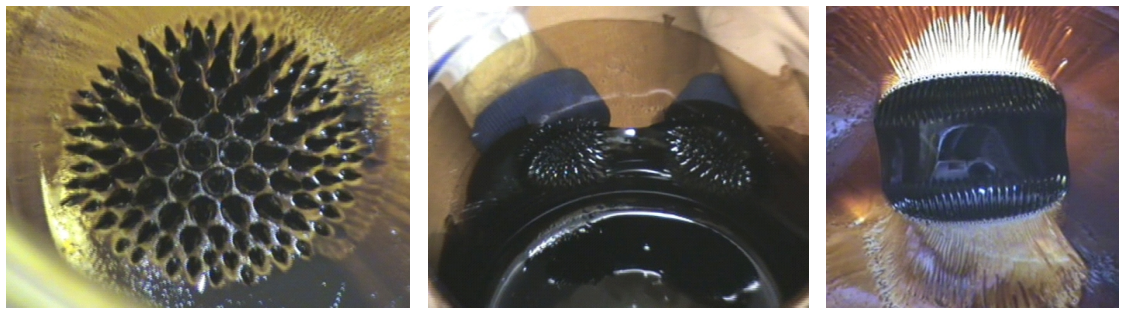

*Figure 8.* Ferrofluid shows lines of magnetic field near a pole, lines between two opposite poles and in case of a dipole.

**Multipoles.** More complex situation can be investigated. Quadrupole (fig. 9) and other multipole configurations can be constructed and studied (fig. 10).

In particular, it is possible to obtain situation with minimum that can be used for discussing how can be obtained a stable configuration in a magnetic train for example.

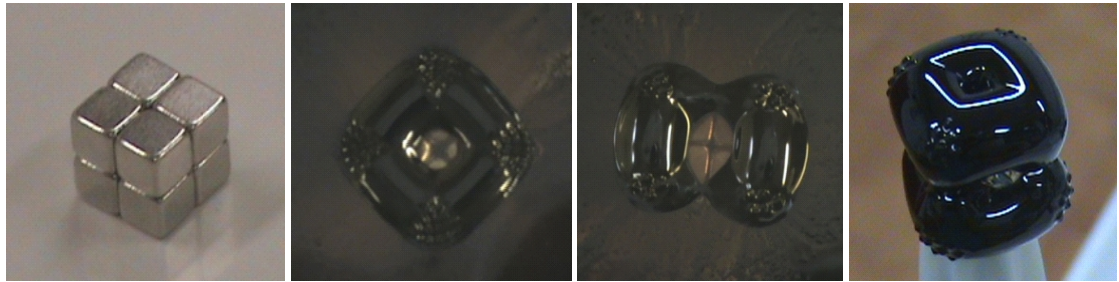

*Figure 9.* Quadrupole and magnetic spatial distribution around it, visualized by a ferrofluid which shows a minimum on the centre of a face.

Another interesting aspect is that student usually have no idea about different field that a multipole can generate so trying to guess the spatial distribution can be a challenging puzzle.

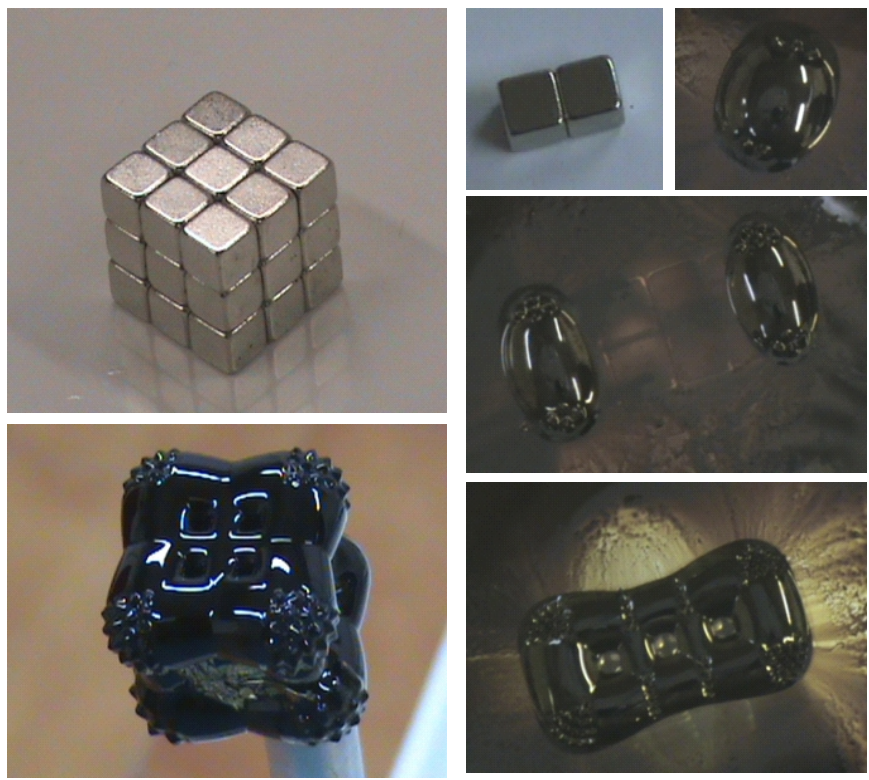

*Figure 10.* A dipole and other multipoles with 3 and 4 minimum point on upper surface and a different behavior on the lateral sides.

Moreover, ceramic magnets can be used for constructing new multipoles easily by assembling many of them, like showed in fig. 9 e 10 where cubic magnets are used. An inquiry-based exploration on multipoles can be proposed by using cubic, spherical and other shape magnets.

**Ferromagnetic behavior** Every situation in which a magnetic field changes or appears or disappears can be useful for a better understanding of magnetic properties of materials. In Fig. 11 a ferromagnetic bolt is examined.

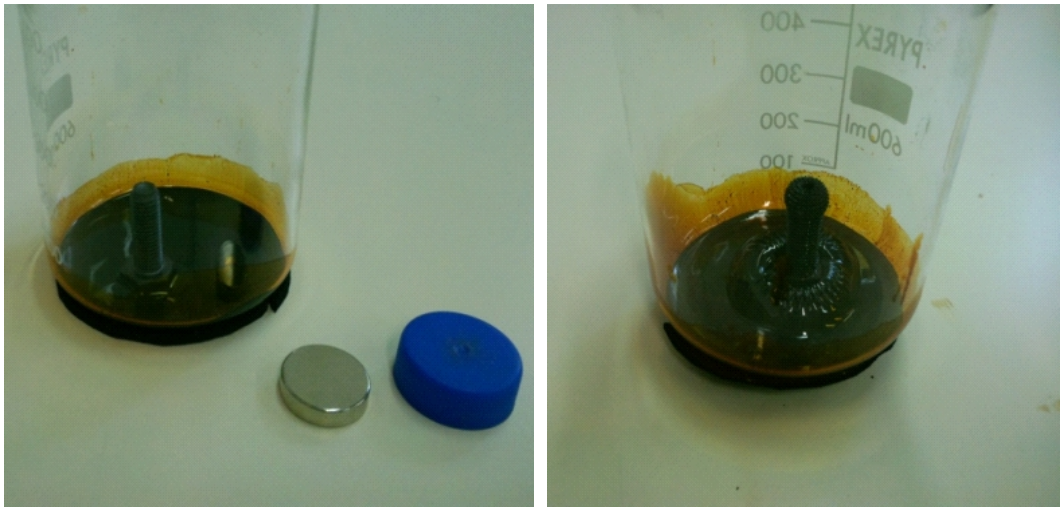

*Figure 11*. A ferromagnetic bolt before magnetization and a massive ceramic magnet, on the left, and after magnetization, obtained by putting the magnet under the becher, on the right. The blue plastic cup and the black rubber interposed between the magnet and the becher are necessary for safety reason, since magnet and ferromagnetic bolt mutually attract strongly.

**Electromagnet.** In this case the solenoid generate no magnetic field until electric current is null. When current increases, ferrofluid allows to see a magnetic field more and more intense as shown in the sequence of pictures in fig. 12.

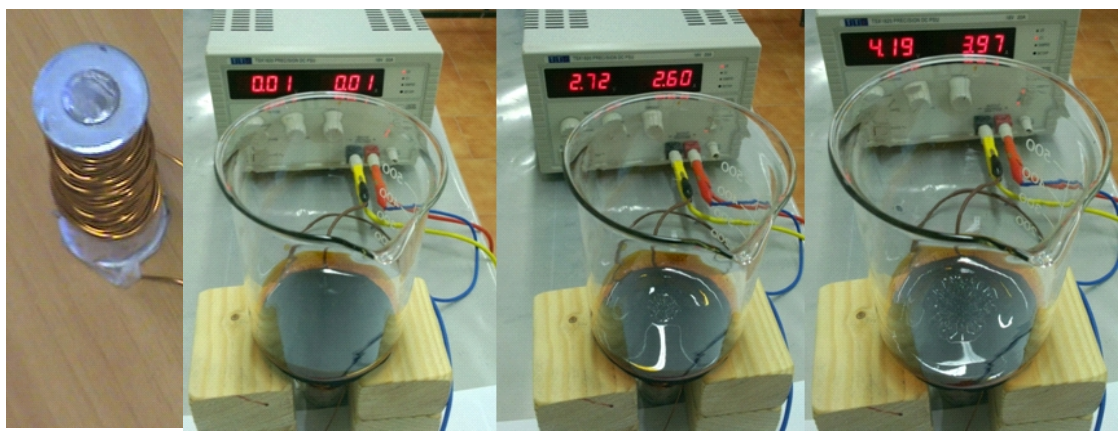

*Figure 12*. An experiment with an electromagnet and ferrofluid. A solenoid (left) is posed between the wooden blocks under the becher containing the ferrofluid. As soon as the electric current increases, the ferrofluid shows the presence of a magnetic fields more and more intense.

### Energy transport and transformations

Energy is one of more important concept in physics but it is always difficult to visualize it. In the following some systems in which energy is transformed and studied is proposed.

*Energy in a Shive wave machine*

Shive wave machine consists of a set of equally spaced horizontal rods attached to a square wire spine. Displacing a rod on one of the ends will cause a wave to propagate across the machine. Torsion waves of the core wire translate into transverse waves.

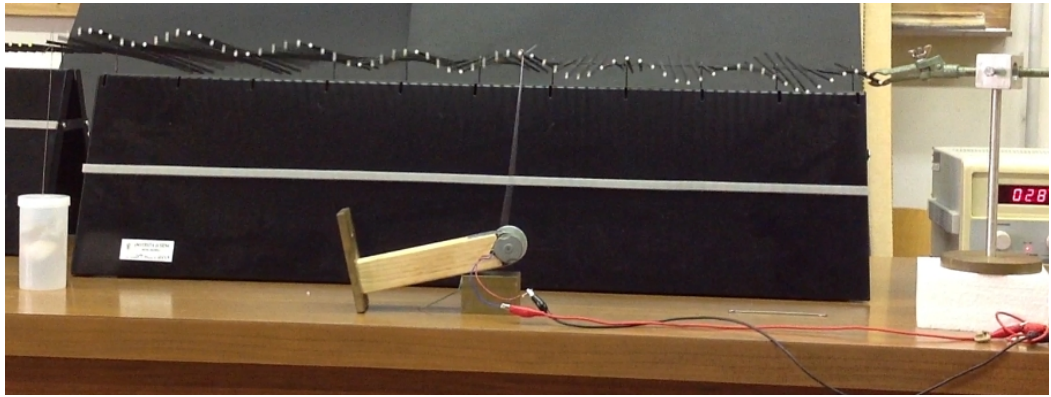

*Figure 13.* A dynamic visualization of transport of energy by a wave.

In fig. 13, an electromechanical system generates a wave at the centre of the Shive machine. The wave propagates from both sides carrying the energy transmitted to the system. On the left of the wave interacts with a mechanical system that dissipates energy into heat, allowing a continuous flow. On the right, the end of the machine is fixed so it behaves as a wall that reflects the wave that overlapping to the incident one creating a standing wave. Shive machine is very effective for encouraging students in active learning (Montalbano & Di Renzone, 2012).

Another example is shown in fig. 14. For example, by coupling core wires of two Shive wave machine with different rod lengths, an impulsive wave can be reflected and transmitted through the discontinuity. Students can measure from captured images all wave amplitudes and speeds and verify that energy is conserved.

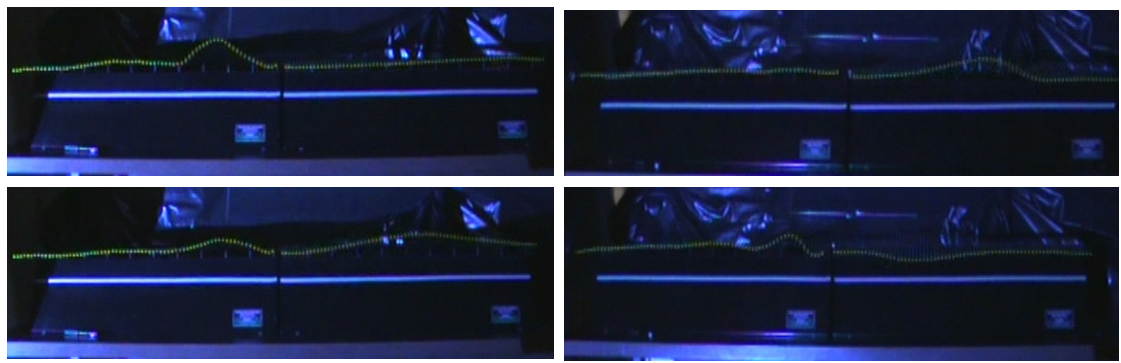

*Figure 14.* On the left, a pulse from the left (top) and transmitted and reflected pulses (below); on the right a pulse from the right (top) and transmitted and reflected pulses. (Montalbano & Di Renzone, 2012).

## *Electromagnetic friction*

Electromagnetic friction can be better understood by using an infrared video camera as shown in fig. 15, where an infra red (IR) image gives thermal distribution in a braking fin of a roller coaster train. Each time that a roller coaster train passes on the braking fin the IR image shows a sudden heating, followed by a slow cooling. In this particular case the energy transformation is very rapid and the video is impressive and can be used for starting a discussion on dissipation in different cases.

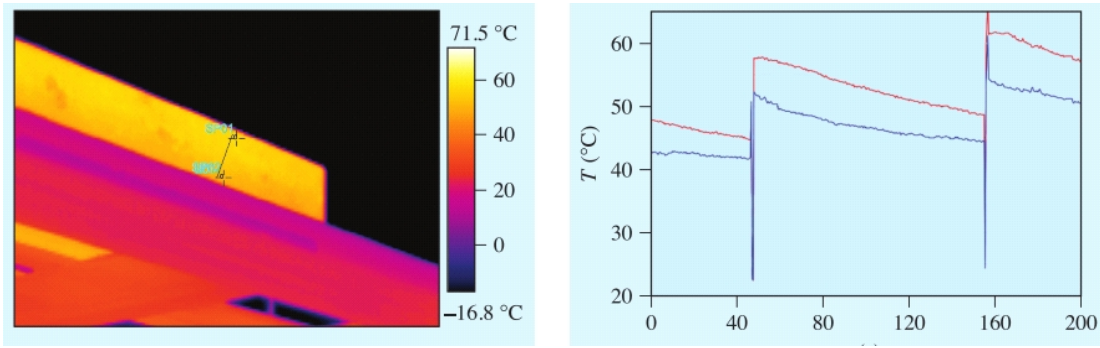

*Figure 15*. On the left, from the digital image it is possible to extract quantitative measurements and time dependence, shown on the right (Pendrill, 2012; Pendrill, Karlsteern, & Rodjegard, 2012).

## RESULTS AND DISCUSSION

Almost all lab activities that can improve external visualization on complex and invisible phenomena were tested in pilot experiences usually with few students in the context of a summer school of physics or in optional laboratory. In order to clarify how visualization in physics lab can improve physics understanding of complex phenomena, the designing of activities in laboratory was focused on the possibility of manipulating, interacting and changing parameters in the system in exam. The results are summarized in Table 2.

Table 2

*Lab activities which allows a direct and interactive visualization were tested in pilot qualitative experiences*

| Lab activity | Context | Participants n. | Participants age | Evaluation |
|---|---|---|---|---|
| Sound waves on oscilloscope screen | optional lab | 7 | 15-16 | effective |
| Magnetic flux | optional lab summer school | 5 + 8 | 16-17 | very effective |
| Magnetic multi poles | proposal | | | |
| Magnetic field spatial distribution | optional lab summer school | 5 + 8 | 16-17 | very effective |
| Ferromagnetic behavior | optional lab summer school | 5 + 8 | 16-17 | very effective |
| Electromagnet | optional lab summer school | 5 + 8 | 16-17 | very effective |
| energy transport | optional lab (pre service teachers) | 6 | 16-17 | very effective |
| Shive machine | | 10 | 25-40 | very effective |
| energy conserv. Shive machine | optional lab | 6 | 16-17 | very effective |
| Electromagnetic friction IR image | proposal | | | |
| Electromagnetic dumping | optional lab summer school | 7 + 10 | 16-18 | very effective |

The last issue in the Table 2 was tested with high school students but it is very interesting like activity to be proposed in a lab for undergraduate students. In fig. 16 the experimental set-up and two interesting function from the oscilloscope are shown. For high school students a measure of dumping time was proposed, but undergraduate student can modeling the physical system and perform a simulation that easily reproduce the function on the screen.

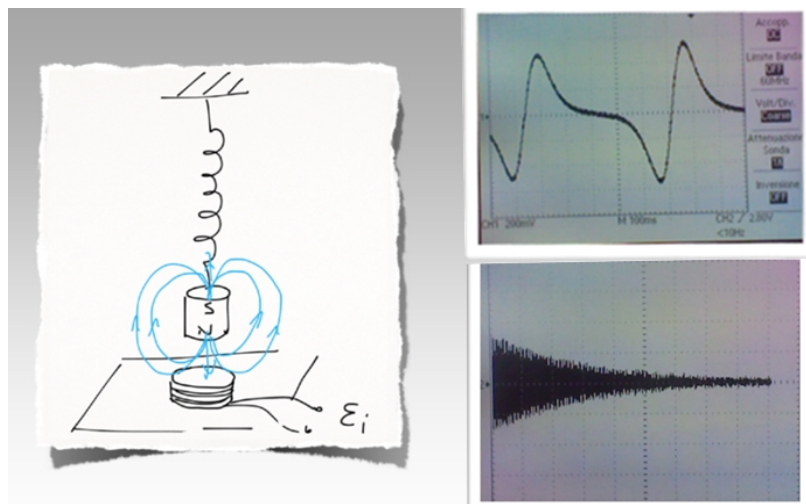

*Figure 16*. Electromagnetic dumping: the experimental set-up on the left, time dependence of the induced voltage across the coil on the right, for short and long time scales.

In conclusions, almost all activities that allow a direct and interactive visualization of complex phenomena seems to be very effective in favouring the learning process.

The next step will be to design learning paths that use these experiences in the lab in order to integrate them in regular teaching in the classroom.

## REFERENCES

AIRES Cosmic Ray Showers website,
    http://astro.uchicago.edu/cosmus/projects/aires/, accessed 2013 November.